# Intensity-Resolved Above Threshold Ionization of Xenon with Short Laser Pulses

N. Hart[1,*], J. Strohaber[1,2], G. Kaya[1], N. Kaya[1], A. A. Kolomenskii[1], H. A. Schuessler[1]

[1]Department of Physics, Texas A&M University, College Station, Texas 77843-4242, USA

[2]Department of Physics, Florida A&M University, Tallahassee, Florida 32307, USA

*Corresponding author: nhart@physics.tamu.edu

*PACS numbers:* 32.80.Rm, 32.80.Fb, 32.90.+a, 42.50.Hz

**Abstract**: We present intensity-resolved above threshold ionization (ATI) spectra of xenon using an intensity scanning and deconvolution technique. Experimental data were obtained with laser pulses of 58 fs and central wavelength of 800 nm from a chirped-pulse amplifier. Applying a deconvolution algorithm, we obtained spectra that have higher contrast and are in excellent agreement with characteristic 2 and 10 $U_p$ cutoff energies contrary to that found for raw data. The retrieved electron ionization probability is consistent with the presence of a second electron from double ionization. This recovered ionization probability is confirmed with a calculation based on the PPT tunneling ionization model [Perelomov, Popov, and Terent'ev, Sov. Phys. JETP **23**, 924 (1966)]. Thus, the measurements of photoelectron yields and the proposed deconvolution technique allowed retrieval of more accurate spectroscopic information from the ATI spectra and ionization probability features that are usually concealed by volume averaging.

# I. INTRODUCTION

The focal volume of a laser beam contains a continuum of intensities that vary both radially and longitudinally with respect to the axis of propagation and range from zero to some peak intensity $I_0$. Each intensity provides a unique ion yield contribution depending on the probability of ionization, $P(I)$, and the volume occupied by the radiation at that intensity. This results in an averaging effect that ultimately reduces the intensity resolution of an experimental measurement [1]. This lack of resolution masks intensity dependent phenomena such as the ionization probability, AC Stark shifts and Rabi oscillations in the atomic energy levels [2]. It has been shown that ions can be distinguished according to their location in the laser focus from which they are produced [3]. But while higher ionization states $A^{+n}$ have been observed in ion time-of-flight (TOF) measurements [1,3], to the best of our knowledge the explicit manifestation of photoelectrons specific to a charge state greater than one has not been observed. The difficulty of such detection follows from the fact that measuring devices are rarely able to determine the location within the focus that an electron originated from. Insufficient temporal resolution results in integration of the signal over the entire focal volume of the laser. For instance, distinguishing two electrons in a field-free region each with 1.5 eV of kinetic energy and a separation distance of 10 μm would require data acquisition electronics with 13 ps temporal resolution. However, fast data acquisition electronics have timing resolutions of a few hundred picoseconds.

Theoretical calculations for laser-matter interactions are typically carried out using plane waves of coherent radiation with some time-dependent amplitude modulation [4,5], and the probability of ionization is determined after the interaction. Because, in practice, ionization experiments with short laser pulses record the ionization yield after the pulse has interacted with the target, and because experimental results are spatially averaged, theoretically determined ionization probabilities are artificially averaged for comparison with experiments. The need to compare with the more fundamental non-spatially averaged theoretical results has motivated the design of intensity-resolved experiments. In them, the goal is to remove the influence of the spatially varying intensity distribution from laser beam modes and isolate the result of a single intensity. Hansch and Van Woerkom [6] used a slit to collect ions from a small cross-section area of the laser focus. The novelty of their approach was that they varied the intensity in which the detected ions were born by changing the position of the slit along the z-axis relative to the laser focus. Walker et. al [7] coupled this measurement with an algorithm that removes the effect of radial variation in the laser intensity. This combined technique is known as Intensity Selective Scanning (ISS). Bryan et. al. [8] modified ISS by accounting for diffraction effects along the z-axis of the laser focus.

Goodworth et. al. [9] developed a deconvolution scheme which used discretized iso-intensity rings of the two dimensional cross-sections of the laser focus. An off axis slit aligned perpendicular to the z-axis determined the width of these cross-sections from which the ions were collected. The volume of each iso-intensity ring was represented by a matrix element $V_{n,s}$ where $n$ indexes the z-axis position and $s$ indexes the intensity of the ring. The deconvolution to obtain a probability $P_s$ was carried out by an inverse matrix $V_{s,n}^{-1}$ multiplication of the yields $Y_n$ from the z-scanned measurement: $P_s = V_{s,n}^{-1} Y_n$.

Other methods have employed purely experimental techniques to measure ions from an iso-intensity volume of the laser focus, which is confined in all three spatial dimensions. Schultze et. al. [10], and Strohaber and Uiterwaal [3] used an imaging TOF spectrometer to sort positive ions from the focus. Ions created at different locations within the focus arrive at a detector at different times. In their experiments, arrival times coupled with longitudinal and transverse measurements provide the ability to both reconstruct the spatial iso-intensity shells of the laser focus and extract intensity-resolved ionization probabilities from intensity scans.

Strohaber et. al. [11] introduced the multiphoton expansion (MPE) as an analytical deconvolution of the laser focal volume by solving the linear Volterra equation of the first kind. The solution for the ionization probability is represented by a power series of the intensity suggesting the name of this approach. The Volterra integral represented the total number of ions detected from an N-dimensional (N = 1, 2 or 3) volume within the focus. As such, this approach allows for the deconvolution of a variety of intensity scanning experimental schemes.

In the present work, we developed a generalized algorithmic technique to recover intensity-resolved above threshold ionization (ATI) energy spectra using photoelectrons; however, the techniques may also be used for other spatially averaged data. It involves obtaining ATI measurements using short laser pulses of different peak intensities and employing a deconvolution algorithm to remove the blurring effect of the spatially varying intensities.

## II. INTENSITY DECONVOLUTION ALGORITHM

The measured ionization yield of an atom $Y(I_0)$ can be expressed as a convolution of the ionization probability per unit volume $P(I)$ and the derivative of the volume $V(I_0, I)$ enclosing all intensities greater than $I$ up to a maximum or peak intensity $I_0$:

$$Y(I_0) = \int_0^{I_0} P(I) \left| \frac{\partial V(I_0, I)}{\partial I} \right| dI , \qquad (1)$$

where $V(I_0, I)$ contains the geometric information about the focal region being measured. Thus, it is implicitly dependent on the optics used and any apertures between the interaction region and the signal detector. The functional form of $V(I_0, I)$ in 1, 2, and 3 dimensions is given in Strohaber et. al. [11].

To deconvolve the ionization probability $P(I)$ from Eq. (1), the experiment must be repeated more than once using different peak intensities. Therefore, we introduce the notation $I_n$ to denote the peak intensity of the laser beam in the $n$-th experiment. We will now construct a numerical approximation of Eq. (1). Note that the magnitude of both $V(I_n, I)$ and its $I$ derivative become infinite as $I$ approaches zero. Therefore the lower limit of Eq. (1) is computationally impractical, and the interval of integration will need to be truncated by a parameter $\delta I \ll I_n - \delta I$:

$$Y(I_n) = \int_0^{\delta I} P(I) \left| \frac{\partial V(I_n, I)}{\partial I} \right| dI + \int_{\delta I}^{I_n} P(I) \left| \frac{\partial V(I_n, I)}{\partial I} \right| dI \approx \int_{\delta I}^{I_n} P(I) \left| \frac{\partial V(I_n, I)}{\partial I} \right| dI. \quad (2)$$

As $I_n$ increases, the integral over the interval $[\delta I, I_n]$ more accurately approximates the full integral over $[0, I_n]$. Since the ionization probability $P(I)$ tends to decay with decreasing intensity, this also reduces the introduced approximation error.

To estimate Eq. (2) numerically, we can discretize the integral using a Riemann sum. The integration interval is partitioned by introducing an ordered set of intensities $I_s \in \{I_1, I_2, ..., I_N\}$ such that $I_1 > I_2 > ... > I_N > \delta I$. Note that our choice of notation for $I_n$ deliberately restricts the set of peak intensities at which we experimentally measure the yield $Y(I_n)$.

Since $V(I_n, I_s)$ monotonically increases with decreasing $I_s$, the volume $V(I_n, \delta I)$ is also implicitly partitioned. We can therefore introduce differential volume elements $\Delta V = V_{n,s}$ for the set $I_s$ at a peak intensity $I_n$ to approximate Eq. (2).

$$\int_{\delta I}^{I_n} P(I) \left| \frac{\partial V(I_n, I)}{\partial I} \right| dI = \lim_{N \to \infty} \sum_{s=n}^{N} V_{n,s} P(I_s) \approx \sum_{s=n}^{N} V_{n,s} P(I_s). \tag{3}$$

$V_{n,s}$ and the corresponding Riemann sum can be defined in a number of different ways (midpoint rule, trapezoidal rule, Simpsons rule, etc…). However, any definition of $V_{n,s}$ must satisfy

$$\sum_{s=n}^{N} V_{n,s} = V(I_n, \delta I), \tag{4}$$

meaning that the sum of all differential volume elements must equal to the total volume enclosed by the smallest intensity $\delta I$. Moreover, to obtain a good approximation to Eq. (3) the condition $V_{n,s} \ll V(I_n, \delta I)$ should be satisfied for all $s$. We chose, for simplicity, to define $V_{n,s}$ by taking the difference between the volumes enclosed by two consecutive iso-intensity shells:

$$V_{n,s} \equiv \begin{cases} |V(I_n, I_{s+1}) - V(I_n, I_s)| \to s \geq n \\ 0 \to s < n \end{cases}. \tag{5}$$

This definition follows from the discrete first derivative of the volume:

$$\frac{\Delta V(I_n, I_s)}{\Delta I} \Delta I = \frac{V(I_n, I_s) - V(I_n, I_{s+1})}{I_s - I_{s+1}} (I_s - I_{s+1}) \tag{6}$$

and is equivalent to taking a right Riemann sum.

If the indices $n$ and $s$ have the same dimensions $(n, s \in \{1, 2, ..., N\})$, Eq. (3) produces a system of linear equations that can be solved for the desired variable $P(I_s)$. Since in this case $I_N$ would be the smallest element in the list of measured intensities, a free parameter $I_{N+1} \equiv \delta I$ must be chosen for the calculation of an outermost volume $V(I_n, I_{N+1})$ for all $n$ (see Fig. 1 and the example below). The determination of $\delta I$ is discussed later at the end of this subsection.

As an example, let us consider a one dimensional case when an experiment is performed at two $(N = 2)$ different laser peak intensities, $I_1 > I_2$, and the volume elements are $V_{1,1}$, $V_{1,2}$ and $V_{2,2}$ (see Fig. 2). Using Eq. (3), the measured ion count rates for beams (a) and (b) in Fig. 2 are then approximated respectively by:

$$Y(I_1) = V_{1,2} P(I_2) + V_{1,1} P(I_1), \tag{7}$$

$$Y(I_2) = V_{2,2} P(I_2). \tag{8}$$

Since the quantities $Y(I_1)$ and $Y(I_2)$ are measured, and $V_{1,1}$, $V_{1,2}$ and $V_{2,2}$ are known from the focal geometry, it is purely a mathematical exercise to solve Eqs. (7) and (8) for $P(I_1)$ and $P(I_2)$:

$$P(I_1) = \frac{1}{V_{1,1}} \left( Y(I_1) - \frac{V_{1,2}}{V_{2,2}} Y(I_2) \right) \tag{9}$$

$$P(I_2) = \frac{Y(I_2)}{V_{2,2}}. \tag{10}$$

For the general case of $N$ different laser peak intensities we can use Eq. (3) to calculate a vector $Y \equiv (Y(I_1), Y(I_2), ..., Y(I_N))^T$ and get a system of linear equations:

$$\begin{pmatrix} V_{1,1} & V_{1,2} & \cdots & V_{1,N} \\ 0 & V_{2,2} & \cdots & V_{2,N} \\ \vdots & \cdots & \ddots & \vdots \\ 0 & 0 & \cdots & V_{N,N} \end{pmatrix} \begin{pmatrix} P(I_1) \\ P(I_2) \\ \vdots \\ P(I_N) \end{pmatrix} = \begin{pmatrix} Y(I_1) \\ Y(I_2) \\ \vdots \\ Y(I_N) \end{pmatrix} \tag{11}$$

or $\hat{V} \cdot P = Y$ where $\hat{V}$ denotes the differential volume matrix, $P$ is the probability array and $Y$ is the signal yield array. To find the probability $P$ we multiply both sides of Eq. (11) by the inverse volume matrix $\hat{V}^{-1}$ to obtain:

$$\boldsymbol{P} = \hat{V}^{-1} \cdot \boldsymbol{Y} . \tag{12}$$

The choice of the free parameter $I_{N+1} = \delta I$ in Eq. (3) can be determined from $V_{N,N} = Y(I_N)/P(I_N)$, which assumes some knowledge of the probability $P(I_N)$. We note that for some simple atoms in the multiphoton regime, the probability $P(I_N)$ can be determined theoretically using perturbation theory [12]. It is known that the multiphoton yield at low intensities is proportional to the probability, since the highest intensity of the beam dominates the signal. We therefore determine $\delta I$ by requiring that the derivatives of the probability and yield are equal at $I_N$, i.e. $\Delta P / \Delta I = \Delta Y / \Delta I_0$.

## III. PROCEDURE FOR REGULARIZATION ALGORITHM

In practice the inversion of Eq. (11) is notoriously unstable, and it is common to remove statistical outliers from the data to improve the algorithm's stability. Here we employ an L2 norm modification of the variation minimization algorithm proposed by Le et. al. [13] and expanded by Chartrand and Wohlberg [14]. Generally, L2 regularization involves the minimization of the dot product $|A|^2$ of a vector $A$, whereas L1 regularization refers to the minimization of the absolute value $|A|$. For convenience of notation we will represent the ionization yields and probabilities in the following way:

$$Y_n \equiv Y(I_n), P_s \equiv P(I_s). \tag{13}$$

From Bayes' theorem, the probability of having a statistical mean $\bar{Y}_n$ given that we measure a yield $Y_n$ can be expressed as:

$$Pr(\bar{Y}_n|Y_n) = \frac{Pr(Y_n|\bar{Y}_n)Pr(\bar{Y}_n)}{Pr(Y_n)}, \tag{14}$$

where $Pr(B)$ is the probability of obtaining $B$ and $Pr(A|B)$ is the probability of obtaining $A$ given that we know $B$. Since $Y_n$, the measurement, cannot be changed, maximizing $Pr(\bar{Y}_n|Y_n)$ requires ascertaining the appropriate $\bar{Y}_n$. Maximizing $Pr(\bar{Y}_n|Y_n)$ is therefore equivalent to maximizing $Pr(Y_n|\bar{Y}_n)Pr(\bar{Y}_n)$. The data $Y_n$ is measured over a fixed interval of time satisfying Poisson statistics. Therefore the probability of measuring $Y_n$, provided a mean $\bar{Y}_n$, is given by the Poisson probability mass function:

$$Pr(Y_n|\bar{Y}_n) = \frac{e^{-\bar{Y}_n}\bar{Y}_n^{Y_n}}{Y_n!}. \tag{15}$$

The regularization of the data is typically introduced through the probability $Pr(\bar{Y}_n)$. However, it is more useful to regularize the output of the deconvolution algorithm $\bar{P}$, since this is where the propagated error tends to be the largest. Consequently, the function $Pr(\bar{Y}_n)$ is replaced by a suitable function $Pr(\bar{P}_n)$. This function must be chosen based upon experimental constraints. Assuming the derivative of the ionization probability to be continuous, we chose:

$$Pr(\bar{P}_n) = exp\left(-\beta\left(\nabla_{n,s}\bar{P}_s\right)^2\right), \tag{16}$$

where the local derivative,

$$\left|\nabla_{n,s}\overline{P}_s\right| \approx \left|\frac{\partial \overline{P}}{\partial I}\right|_n, \tag{17}$$

is with respect to the array variable $I$, and $\beta$ is the regularization parameter. The choice of $\beta$ is discussed in Section V. Since we ultimately seek the statistical mean of the ionization probability $\overline{P}$, we eliminate the yield mean by the substitution

$$\overline{Y} \to \hat{V}\overline{P}. \tag{18}$$

We can now maximize Eq. (14) by minimizing its negative logarithm:

$$\begin{aligned}-\log\left(Pr\left(Y_n|\overline{Y}_n\right)Pr\left(\overline{P}_n\right)\right) &= \overline{Y}_n - Y_n \log\left(\overline{Y}_n\right) + \log\left(Y_n!\right) + \beta\left(\overline{Y}_n\right)^2 \\ &= V_{n,s}\overline{P}_s - Y_n \log\left(V_{n,s}\overline{P}_s\right) + \log\left(Y_n!\right) + \beta\left(\nabla_{n,s}\overline{P}_s\right)^2. \end{aligned} \tag{19}$$

Equation (19) can be viewed as a mechanical action, from which we derive the Euler-Lagrange equation with respect to the variables $\overline{P}_s$ and $\nabla_{n,s}\overline{P}_s$, resulting in

$$\hat{V}^T \frac{\hat{V}\overline{P} - Y}{|\hat{V}\overline{P}|} - 2\beta\hat{\nabla}^2\overline{P} = 0. \tag{20}$$

It should be noted that in Eq. (20) $\hat{M} \equiv |\hat{V}\overline{P}|$ is a diagonal matrix whose elements are:

$$M_{n,m} = \begin{cases} V_{n,s}\overline{P}_s \to n = m \\ 0 \to n \neq m \end{cases}. \tag{21}$$

This $\hat{M}$ has the general effect of rescaling the regularization parameter $\beta(I_n) = \beta M_{n,n}$ to accommodate the variation in the Poisson noise. Because $\hat{M}$ is a function of $\overline{P}$ (and $\overline{P}$ is the desired quantity), $\hat{M}$ will have to be approximated through an iterative process. If the experimental data is taken such that the measurement approximates the statistical mean, $Y \approx \bar{Y} = \hat{V}\bar{P}$, we can approximate Eq. (21) by setting the initial value $\hat{M}_0 = |Y|$ and solving for the probability $\bar{P}_i$:

$$\bar{P}_i = \left(\hat{V}^T\hat{V} - 2\beta\hat{M}_i\hat{\nabla}^2\right)^{-1} \hat{V}^T Y, \tag{22}$$

$$\hat{M}_{i+1} = |\hat{V}\bar{P}_i|. \tag{23}$$

Equations (22) and (23) are iterated until convergence ($|\bar{P}_i - \bar{P}_{i+1}| < |\bar{P}_{i+1}|/10^4$) is obtained for every element of the vector $\bar{P}_{i+1}$. For our data, only two iterations were needed for convergence. In Eq. (22), the initial $\hat{M}_i$ ($i = 0$) is the diagonal matrix of the measured yields

$$|Y| = \begin{pmatrix} |Y_1| & 0 & \cdots & 0 \\ 0 & |Y_2| & 0 & \vdots \\ \vdots & 0 & \ddots & 0 \\ 0 & 0 & 0 & |Y_n| \end{pmatrix}, \tag{24}$$

and $\hat{\nabla}^2$ is the second derivative matrix. We found the second derivative by multiplying two first derivative matrices defined by

$$\hat{\nabla} = \frac{1}{\Delta I}\begin{pmatrix} 1 & -1 & 0 & . & 0 \\ 0 & 1 & -1 & 0 & . \\ . & 0 & 1 & . & 0 \\ . & . & 0 & . & -1 \\ 0 & . & . & 0 & 1 \end{pmatrix}. \tag{25}$$

In general, the intensity spacing $\Delta I = I_i - I_{i+1}$ is not constant and should be calculated for each row of the derivative matrix. The initialization step along with the iteration of Eqs. (22, 23) and the convergence criterion will hereafter be referred to as the discrete deconvolution and regularization (DDAR) algorithm.

## IV. EXPERIMENTAL SETUP

The ATI apparatus is depicted in Fig. 3. Target xenon atoms were ionized with short laser pulses. A series of ionization measurements was taken for 120 different peak laser intensities ranging within $3 \times 10^{13} - 8 \times 10^{14} \, \text{W/cm}^2$. All other laser parameters, such as mode quality, pulse duration and spectral bandwidth were unchanged.

The Ti:Sapphire laser oscillator provides 20 fs mode-locked laser pulses at a repetition rate of 80 MHz. These pulses are seeded into a regenerative laser amplifier, which outputs 58 fs (measured by frequency resolved optical gating, GRENOUILLE 8-20, Swamp Optics, LLC) laser pulses at a repetition rate of 1 kHz, and a central wavelength of 800 nm. Since shorter pulses have a higher peak intensity for a given pulse energy, temporal compression of the laser pulses in the focus was achieved by maximizing the integrated photoelectron yield in the ATI apparatus by adjusting the grating compressor in the laser amplifier. The maximum pulse energy was approximately 0.8 mJ.

Laser pulses were detected before the half-wave plate of the attenuator by a photodiode, and the signal was used to trigger the data acquisition software. The attenuator consisted of a half wave plate that changed the polarization of the initially horizontally polarized light and a polarizing cube that filtered out vertically polarized light, while horizontally polarized light passed through. The orientation of the wave plate was chosen such that after the polarizing cube the desired intensity is achieved in the laser focus.

The vacuum chamber was filled with xenon gas of 99.999% purity (Advanced Specialty Gasses) through a variable leak valve. The xenon pressure ($5\times10^{-6}$ mbar) was three orders of magnitude higher than the background pressure in the ionization chamber. Because the ionization potential of water (12.61 eV) is roughly equal to that of xenon (12.15 eV), a large-surface-area vacuum feed-through, located on a remote region of the time-of-flight (TOF) chamber, was chilled using liquid nitrogen to freeze out the residual water molecules from the background vacuum. The laser beam was focused by a 20 cm achromatic lens. Ionized electrons were ejected along the polarization of the laser field in the direction of the microchannel plate (MCP) detector. The electrons travelled within a µ-metal TOF tube in a field-free region. Electrons from the entire focal volume of the laser were measured at the detector. The signals from the MCP were amplified by a high bandwidth Mini-Circuits ZKL-2 pre-amplifier before being registered by a FAST ComTec MCS6 multiscaler with 100 ps timing resolution. A power meter (PM) measured the average laser power, which is proportional to the average peak laser intensity in the focus.

The DDAR algorithm was written in Mathematica® and employed on an Intel i7 desktop computer having 16 GB of memory. The algorithm deconvolved the entire data set (a 19.0MB matrix of raw electron TOF spectra) in 0.824s.

## V. RESULTS

The electron ionization yield as recorded along the laser polarization is shown in Fig. 4. The saturation intensity is measured to be $I_{sat} = 1.2\times10^{14}\,\text{W/cm}^2$. On a Log-Log plot the yield curve shows a slope of 5 for intensities less than $I_{sat}$ and a slope of 3/2 for intensities greater than $I_{sat}$. The slope of 3/2 arises from volumetric integration of the electrons ionized from all intensities in the Gaussian beam. As the peak intensity $I_0 \equiv I(r=0)$ increases, the total volume enclosed by an intensity $I(r>0) < I_0$ grows as $I_0^{3/2}$ [15]. As this volume grows, so does the yield. However, the largest contribution to the yield after the saturation intensity come from those intensities with the highest ionization probability.

One of the effects of using regularization is that the resulting yield $\bar{Y}$ is smoother than the original data. This provides more stability to the retrieved probability $\bar{P}$. Increasing the regularization parameter $\beta$ strengthens the regularization and minimizes discontinuities in the derivative of $\bar{P}$. Consequently, we used $\beta = 0.5$. Since $\bar{P}$ is the ionization probability per unit volume, we divide it by the gas density (proportional to pressure) in the laser interaction region to obtain the ionization probability per atom. Electrons from different ion charge states have unique ionization probability functions that approach unity as intensity increases. However, these charge states have different saturation intensities. Hence, the graph of the probability first saturates (approaches 1) at $1.2 \times 10^{14}$ W/cm$^2$ and then reaches a maximum value of 2 at approximately $2.7 \times 10^{14}$ W/cm$^2$ (Fig. 5). This second saturation is attributed to double ionization. The MCP detector cannot distinguish between electrons from different charge states. Therefore, electron yields from both species and, by implication, their probabilities are summed giving a "stair step" appearance. In addition to the deconvolved experimental data, Fig. 5 also shows the results of a Perelomov, Popov and Terent'ev (PPT) tunneling ionization simulation [16]. The red curve is the result of summing the calculated ionization probability of both the Xe$^+$ and Xe$^{+2}$ ions, whereas the blue curve exclusively represents the Xe$^{+2}$ ionization probability.

Even though the signal shows significant noise, DDAR still recovers the probability. The second ionization of Xe has been measured by other groups using ion but not electron detection as in this experiment and compares favorably with our results [1]. The counting electronics naturally groups the electrons according to when they arrive or their TOF. By transforming this time-series into an energy spectrum and applying Eq. (22) to the yield rates for each electron energy the intensity-resolved (volume independent) energy spectra are obtained. One such spectrum is plotted in Fig. 6.

For the following discussion of features in the ATI spectra see Fig. 6. The first plateau between 0 and 8 eV is the result of "direct" electrons that do not scatter off the parent ion after being ionized. These electrons have a classical cutoff energy of $2U_p$, where $U_p$ is the ponderomotive energy of the laser field [17]. In this low energy region, resonantly enhanced multi-photon ionization (REMPI) is expected to dominate the ATI peak structure (inset) [18]. The second plateau between 12 and 25 eV is dominated by the interference of electrons which follow different quantum trajectories and are freed with an initially near zero velocity by either tunneling ionization or resonant multiphoton ionization at a channel closing [19]. This ionization mechanism makes the second plateau important also for the study of high harmonic generation [21]. As electrons are accelerated by the electric field, the different quantum paths of electrons with equal momenta can constructively interfere with each other leading to an enhancement in the ionization yield [19,20]. The third plateau, which ranges from 30 to 50 eV, corresponds to elastic backscattering of the electron off the parent ion. This plateau has a cutoff energy of $10.007U_p$ due to the maximum classical energy that a backscattered electron can have [17].

In Fig. 6, the experimental data shows a $10U_p$ value of $\approx 40$ eV, which is smaller than that of the deconvolution ($\approx 45$ eV). This can be explained by the fact that the peak intensity for each of our Gaussian beams has the smallest three dimensional volume. In our case, we can verify this explicitly by calculating the volume elements of the beams at each peak intensity (Eq.

(5)). Figure 7 shows density plots of the ATI spectra as a function of the electron energy (horizontal axis) and laser intensity (vertical axis). The dotted curves drawn on top of the density plots are the $2U_p$ and $10U_p$ cutoff energies calculated from the formula:

$$U_p[\text{eV}] = 9.33 \times 10^{-14} I_0[\text{W/cm}^2] \lambda^2 [\mu\text{m}^2], \tag{26}$$

where $\lambda$ is the center wavelength in micrometers, $I_0$ is the intensity in $\text{W/cm}^2$ and the resulting ponderomotive energy has units of eV. For our raw experimental data (Fig. 7(a)), the measured $2U_p$ and $10U_p$ values for each intensity were smaller than values calculated with the DDAR algorithm. This discrepancy could not be removed by adjusting the intensity calibration by a scaling factor. The deconvolution however gives good agreement with the calculated cutoff energies (Fig. 7(b)). So even though the ionization probability is in general higher for larger intensities, the ionization contributions from intensities slightly lower than the peak value can significantly change the spectrum due to their larger volumes. This is important, because it means that the peak intensity and energy of a laser pulse cannot be calculated directly from volume integrated data using the cutoff energies of the spectrum. It should also be noted that none of the spectra from the set of raw data show as much contrast in the ATI peaks as the deconvoluted energy spectra.

## VI. CONCLUSION

The volume integration in the laser focus reduces the intensity resolution of an experimental measurement. Therefore, we developed a discrete deconvolution and regularization (DDAR) algorithm and applied it to the xenon photoelectron yield to obtain ionization probabilities and intensity-resolved ATI spectra. Our results show that both single and double ionization probabilities is observed by inverting the electron yield with DDAR. The retrieved $Xe^+$ ATI spectrum showed sharper peaks throughout the entire energy range compared to the directly measured one. The $2U_p$ plateau region, where femtosecond pulse ionization from Rydberg states is known to dominate the spectrum, also shows increased contrast after application of the algorithm.

Applying DDAR also recovered $2U_p$ and $10U_p$ cutoff energies that are in excellent agreement with theory while the experimental data is not. In the latter, intensities that are below the peak intensity can dominate the ATI spectrum due to their much larger differential volumes. Consequently, it leads to a discrepancy between the intensity predicted from the $10U_p$ cutoff energy and the actual peak intensity. This discrepancy cannot be removed by rescaling the intensity calibration by a multiplicative factor. Therefore, we found that the volume averaging effect can lead to an underestimation of the $10U_p$ cutoff energy (and this discrepancy grows with increasing intensity) by as much as 30%.

## VII. ACKNOWLEDGMENTS

This work was funded by the Robert A. Welch Foundation Grant No. A1546 and the Qatar Foundation under the grants NPRP 5 - 994 - 1 – 172 and NPRP 6 - 465 - 1 - 091.

REFERENCES

[1] S. Larochelle, A. Talebpour, and S. L. Chin, J. Phys. B: At. Mol. Opt. Phys. **31**, 1201 (1998)

[2] P. Lambropoulos, AIP Conf. Proc., **275**, 499 (1993).

[3] J. Strohaber and C. Uiterwaal, Phys. Rev. Lett. **100**, 023002 (2008).

[4] K. J. Schafer and K. C. Kulander, Phys. Rev. A, **42**, 5794 (1990).

[5] E. Cormier and P. Lambropoulos, J. Phys. B, **30**, 77 (1997).

[6] P. Hansch and L. D. Van Woerkom, Opt. Lett., **21**, 1286 (1996).

[7] M. A.Walker, L. D. Van Woerkom, and P. Hansch, Phys. Rev. A, **57**, R701 (1998).

[8] W. A. Bryan, S. L. Stebbings, E. M. English, T. R. Goodworth, W. R. Newell, J. McKenna, et al. Phys. Rev. A, **73**, 013407 (2006).

[9] T. R. J. Goodworth, W. A. Bryan, I. D. Williams, and W. R. Newell, J. Phys. B **38**, 3083 (2005).

[10] M. Schultze, B. Bergues1, H. Schröder1, F. Krausz, and K. L. Kompa, New J. Phys., **13**, 033001 (2011).

[11] J. Strohaber, A. A. Kolomenskii, H. A. Schuessler, Phys. Rev. A, **82**, 013403 (2010).

[12] R. W. Boyd, *Nonlinear Optics* (Academic Press, Waltham, MA, 2008).

[13] T. Le, R. Chartrand and T. Asaki, JMIV, **27**, 257 (2007).

[14] R. Chartrand and Wohlberg, B., Presented at the IEEE ICASSP, Dallas, TX, USA, (2010).

[15] S. Speiser and J. Jortner, Chem. Phys. Lett. **44**, 399 (1976).

[16] A. M. Perelomov, V. S. Popov, and M. V. Terent'ev, Zh. Eksp. Teor. Fiz. **50**, 1393 (1966) [Sov. Phys. JETP **23**, 924 (1966)].

[17] G. G. Paulus, W. Becker, W. Nicklich, H. Walther, J. Phys. B, **27**, L703 (1994).

[18] McIlrath, T.J.,Freeman, R.R., Cooke, W. E. and Woerkom, L.D. Phys. Rev. A **40**, 2770. (1989).

[19] G. G. Paulus, F. Grasbon, H. Walther, R. Kopold, and W. Becker, Phys. Rev. A, **64**, 021401(R). (2001).

[20] G. G. Paulus, F. Grasbon, A. Dreischuh, H. Walther, R. Kopold and W. Becker Phys. Rev. Lett. **84**, 3791–3794 (2000).

[21] R. Kopold, W. Becker, M. Kleber and G. G. Paulus, J. Phys. B: At. Mol. Opt. Phys. **35** 217 (2002).

FIGURE CAPTIONS

FIG. 1. (Color outline) An example schematic in one dimension showing how volume elements are related to peak intensities. Here the total number of experiments is $N=3$. The boundary of each volume (horizontal) is set by intensities $\delta I$ and $I_3$. $\delta I$ is a free parameter that provides an outer boundary for the calculation of the volume elements. The central blue region represents $V_{3,3}$, the sandwiched red regions are $V_{2,3}$ and the outer gold regions are $V_{1,2}$. Each of these three volumes correspond to the same ionization probability $P(I_3)$.

FIG. 2. (Color outline) A one dimensional illustration of Gaussian beams showing the relationship between the volume elements ($V_{1,1}$, $V_{1,2}$, $V_{2,2}$) and their respective probabilities ($P(I_1)$, $P(I_2)$). Regions within the beams with an ionization probability of $P(I_1)$ are colored blue, while regions with probability $P(I_2)$ are colored red. Beam (a) is represented by Eq. (7). The differential volume of the red region is denoted by $V_{2,2}$. Beam (b) is represented by Eq. (6). Here the differential volume of the red region is denoted by $V_{1,2}$ and that of the blue region is denoted by $V_{1,1}$. The gold wings of each beam are neglected in Eq. (3) and (4).

FIG. 3. (Color outline) Experimental setup: M, mirror; WP, half wave plate; PD, photo diode; PBC, polarizing beam-splitter cube; L, lens; MCP, chevron microchannel plate; PM, power meter.

FIG. 4. The experimentally measured electron yield **Y** on a Log-Log plot. Two slopes are plotted showing the intensity dependence: a slope of 5 (solid line), and a slope of 3/2 (dashed line). The change in slope occurs slightly above the saturation intensity $I_{sat} = 1.2 \times 10^{14}$ W/cm$^2$.

FIG. 5. Recovered electron probability on a Log-Log plot (dotted line). The red curve is a PPT theory simulation of the xenon probability for electron yields for Xe$^{+1}$ and Xe$^{+2}$. The blue curve shows the simulation of the electron yield probability for Xe$^{+2}$ electrons alone. The deconvolution diverges at the high intensity end point.

FIG. 6. Intensity-resolved ATI energy spectra at $8.7 \times 10^{13}$ W/cm$^2$ of electrons per laser pulse. The red curve is the measured data prior to being deconvolved. The deconvolution shows more pronounced features. The inset shows the $2U_p$ low energy region of the same data. REMPI peaks can be seen at energies less than 4 eV.

FIG. 7. Density plots of the ATI spectra as a function of energy (horizontal-axis) and intensity (vertical-axis). In both graphs the $2U_p$ and $10U_p$ cutoff energies for each intensity are denoted by the dotted black and white lines, respectively. (a) The density plot of the experimental data shows a discrepency between the calculated cutoff energies and the measured ones. (b) The deconvolution of the experimental data recovers the calculated cutoff energies and suggests a better agreement with theory [17].

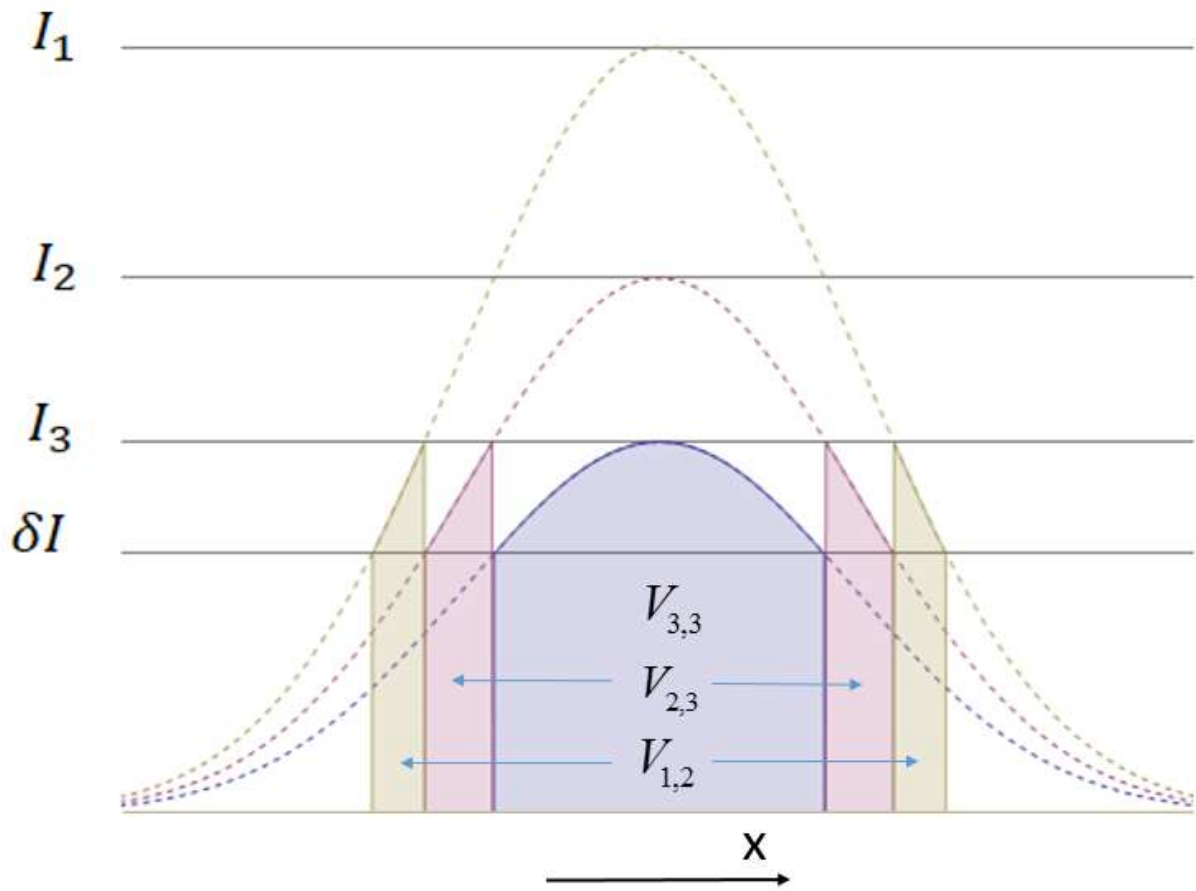

Fig. 2

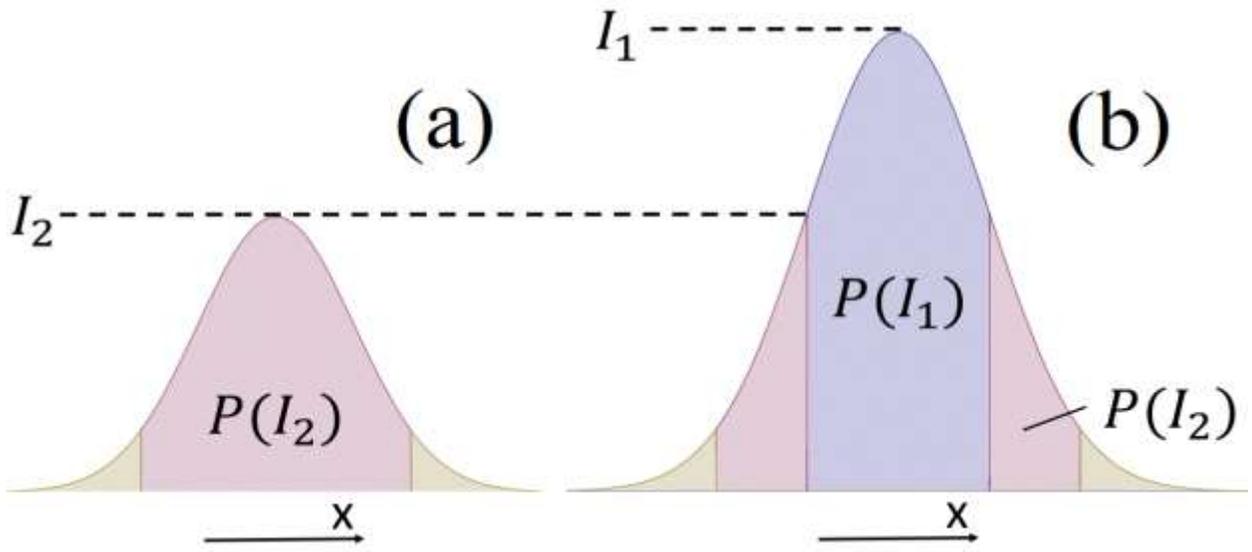

Fig. 3

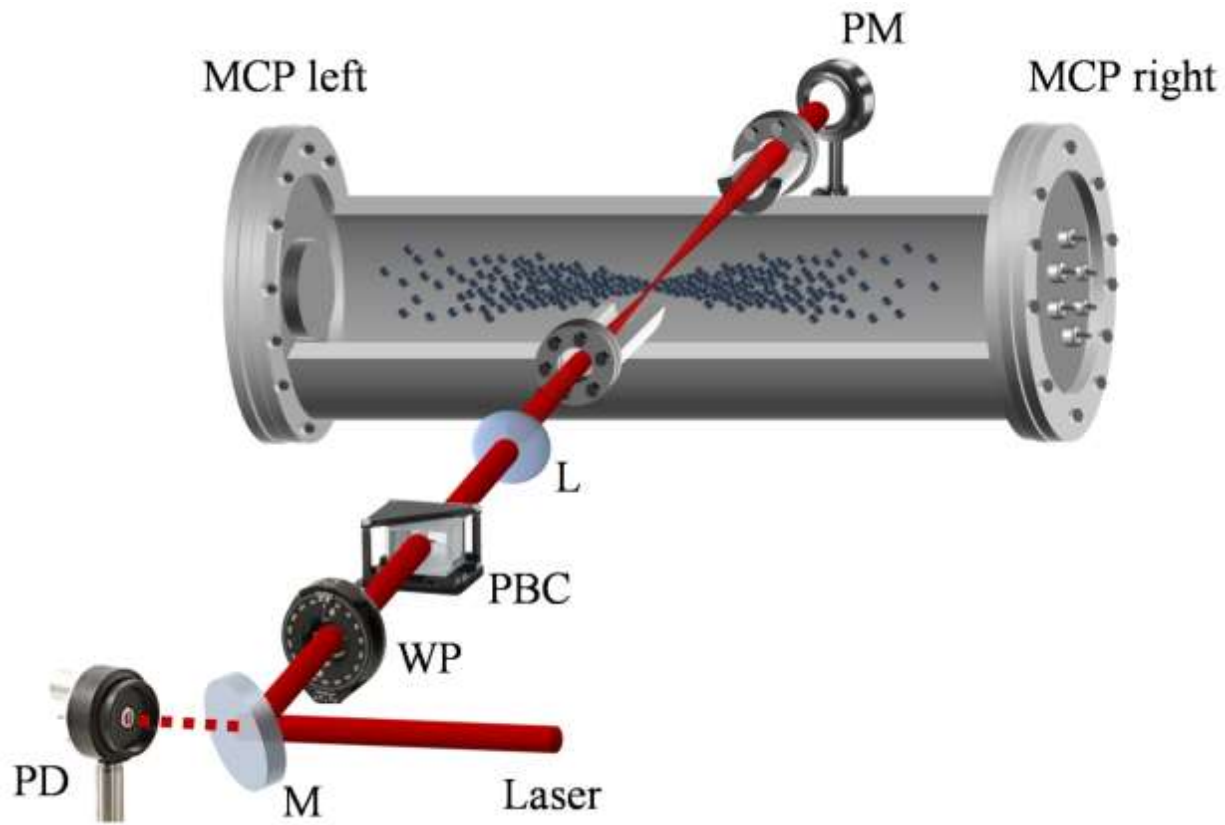

Fig. 4

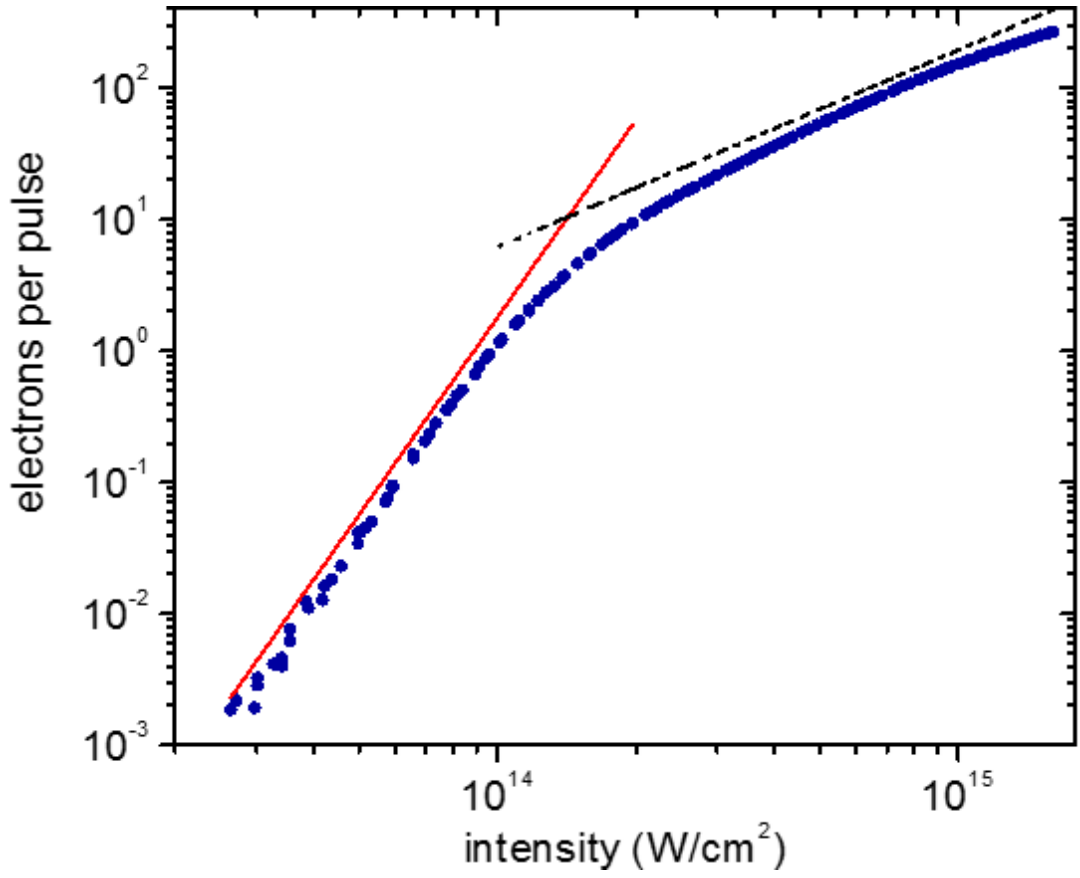

Fig. 5

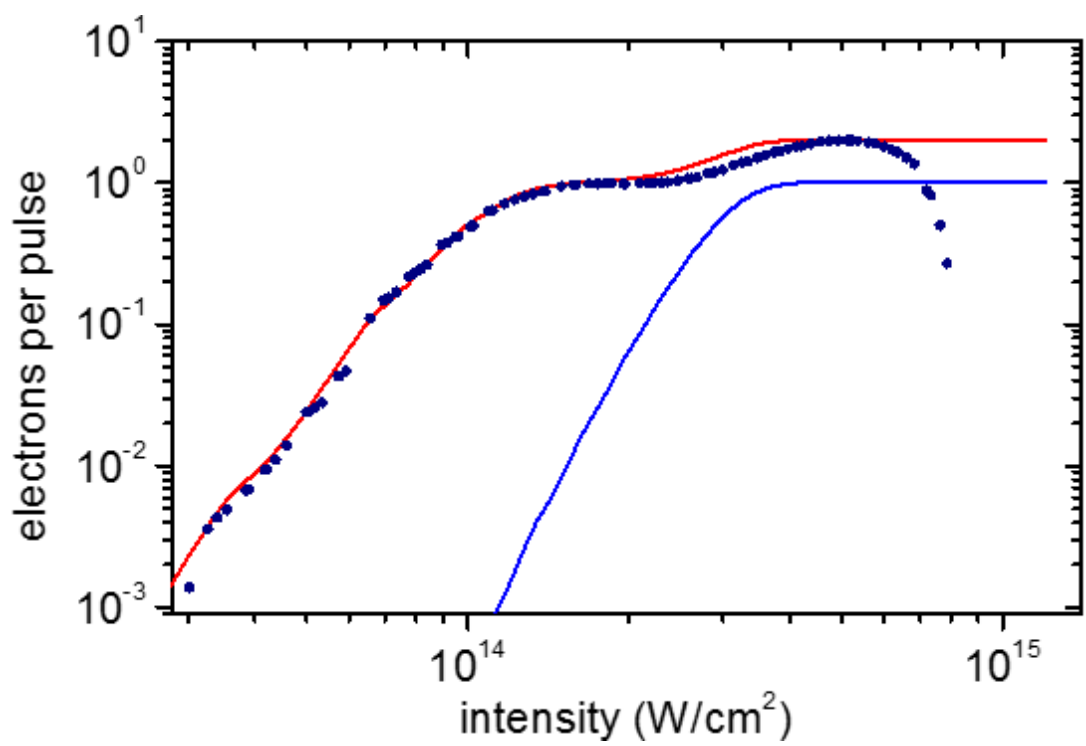

Fig. 6

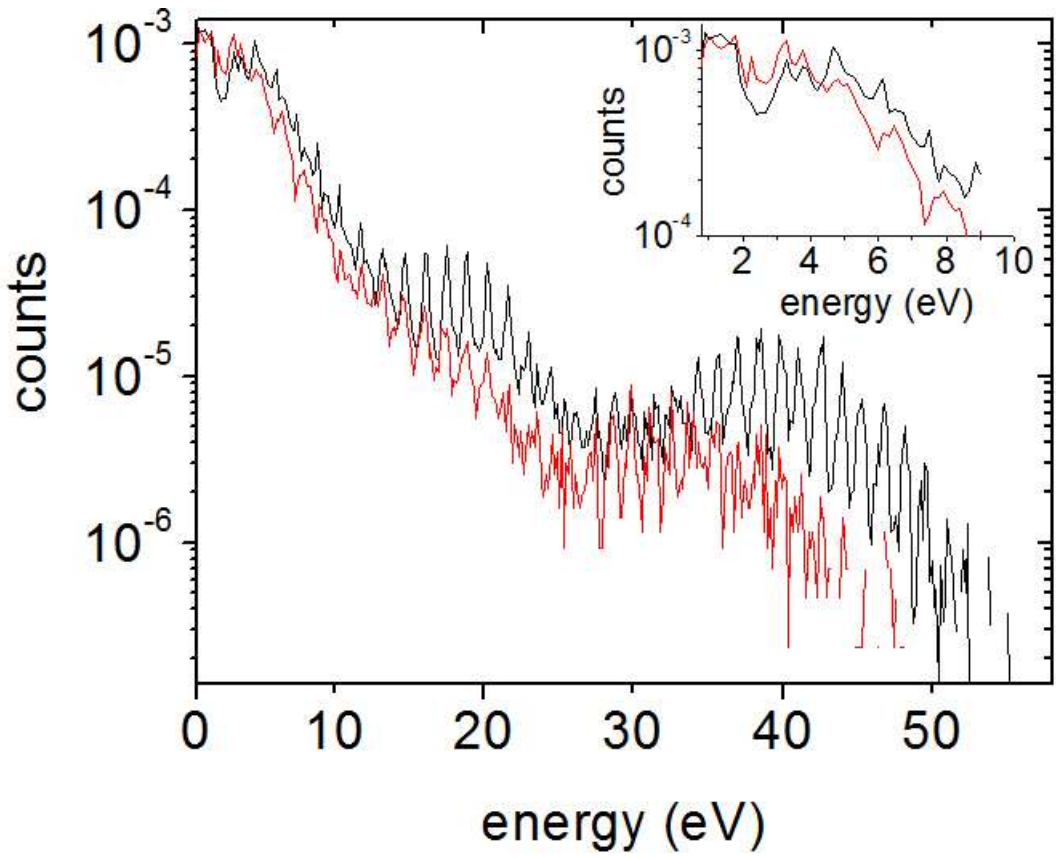

Fig. 7

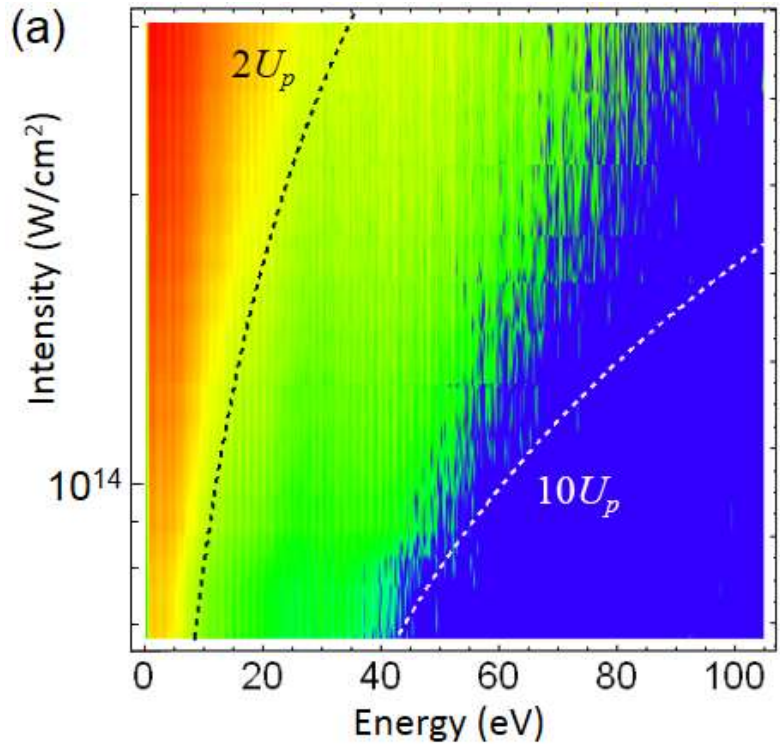

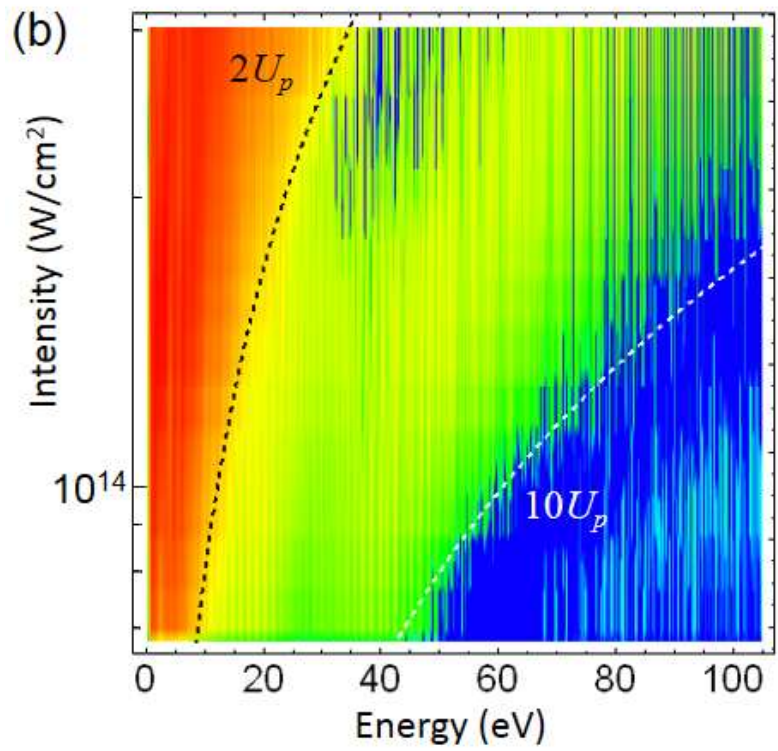